\documentclass[pra,aps,twocolumn,amsmath,amssymb]{revtex4}

\usepackage{graphicx}% Include figure files
\usepackage{dcolumn}% Align table columns on decimal point
\usepackage{bm}% bold math
\usepackage[extension=xxx]{hyperref}

\newcommand{\beq}{\begin{equation}}
\newcommand{\eeq}{\end{equation}}
\newcommand{\beqa}{\begin{eqnarray}}
\newcommand{\eeqa}{\end{eqnarray}}

\def\beq{\begin{equation}}

\usepackage{color}

\begin{document}

\title{Out-of-equilibrium dynamics of a Bose Einstein condensate in a periodically driven band system}
\date{\today}

\author{E. Michon$^1$, C. Cabrera-Guti\'errez$^1$, A. Fortun$^1$, M. Berger$^1$, M. Arnal$^1$, V. Brunaud$^1$, J. Billy$^1$, C. Petitjean$^2$, P. Schlagheck$^2$, D. Gu\'ery-Odelin$^1$}

\affiliation{$^1$ Universit\'e de Toulouse ; UPS ; Laboratoire Collisions Agr\'egats R\'eactivit\'e, IRSAMC ; F-31062 Toulouse, France} 
\affiliation{CNRS ; UMR 5589 ; F-31062 Toulouse, France}
\affiliation{$^2$ D\'epartement de Physique, CESAM research unit, University of Liege, 4000 Li\`ege, Belgium} 

\begin{abstract}
We report on the out-of-equilibrium dynamics of a Bose-Einstein condensate (BEC) placed in an optical lattice whose phase is suddenly modulated. The frequency and the amplitude of modulation are chosen to ensure a negative renormalized tunneling rate. Under these conditions, staggered states are nucleated by a spontaneous four wave mixing mechanism. The nucleation time is experimentally studied as a function of the renormalized tunnel rate, the atomic density and the modulation frequency. Our results are quantitatively well accounted for by a Truncated Wigner approach and reveal the nucleation of gap solitons after the quench. We discuss the role of quantum versus thermal fluctuations in the nucleation process and experimentally address the limit of the effective Hamiltonian approach.

\end{abstract}
\maketitle

Cold atoms provide powerful and versatile platforms for quantum simulators of many-body systems \cite{RMPStringariBoson,RMPStringariFermion,RMPDalibard1,BlochNaturePhysics}, and give access to the rich out-of-equilibrium dynamics of such systems. A remarkable progress for tunability has been achieved with the renormalization of the tunneling rate in time-dependent double-well potentials and optical lattices with the periodic shaking of the potential energy landscape \cite{ChuPRL,ArimondoPRL,Oberthaler}. It has opened many new possibilities for quantum simulations with the possibility to engineer effective Hamiltonians and simulate topological phases \cite{PRXJean}. This includes the realization of the Hofstadter  \cite{Hofstadter} and Haldane models \cite{Haldane} and the investigation of frustrated magnetism \cite{sengstock}, to name a few recent examples.

\begin{figure}[t]
\centering
\includegraphics[width=8cm]{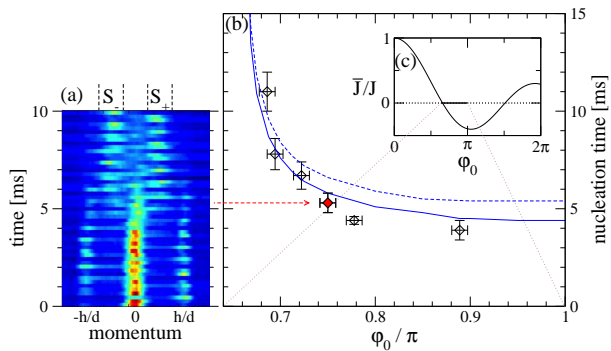}
\caption{(Color online) (a) Transition to a staggered state in the presence of a periodically shaken optical lattice with the depth $V_0 = 2.6 E_L$, the driving frequency $\nu = 1.5$ kHz, and the shaking amplitude $\varphi_0 = 0.75\pi$ for which the effective time-averaged inter-site tunneling matrix element $\bar{J}$ becomes negative [see Eq.~\eqref{jeff} and inset (c)]. The absorption images obtained for various evolution times (followed by $25$ ms time of free flight in the absence of the lattice) clearly display the passage of a spatially periodic condensate wavefunction to the population of a staggered state lying at the edge of the Brillouin zone.
(b) Nucleation time of the formation of staggered states for various values of the shaking amplitude $\varphi_0$ for which the effective tunneling matrix element is negative, as shown in the inset (c). The solid (dashed) line corresponds to the numerical results obtained using the TW approach and assuming the presence of $10^5$ ($5\times 10^4$) atoms within the condensate. A diverging nucleation time is obtained for $\varphi_0=0.6588\pi$ at which the first zero of the Bessel function arises in (c).}
\label{figj0}
\end{figure}

This renormalization can be readily understood in the context of a one-body analysis. However, interactions play a key role in the quantum transition triggered by the modulation. An interesting situation in this respect is provided by the phase modulation of a 1D optical lattice in a regime for which the renormalized tunneling rate becomes negative. In this regime, the quantum gas experiences  a dynamical instability associated with a spontaneous four wave mixing mechanism \cite{ChuPRL,molder,ketterle}. This generates a new quantum state, commonly called a staggered state, for which neighbouring sites acquire opposite phases \cite{ChuPRL,ArimondoPRL}. This phase transition can be readily observed in momentum space after a time of flight expansion. The interference patterns \cite{inguscio}  observed as a result of the spatial periodicity of a static optical lattice are modified: new peaks in between the former static peaks are observed [see Fig.~\ref{figj0}(a)].  Interestingly this transition is not quantitatively accounted for with a standard mean-field approach. 

In this article, we experimentally investigate the nucleation of such staggered states and compare our result with a time-dependent beyond mean-field approach. Our experiments are complementary to nucleation studies of vortices in a BEC using the rotating spoon technique \cite{jeanvortex1,foot,cornell}, which provides another example of phase transition triggered by a dynamical instability \cite{yvan1,yvan2}. In those latter experiments, however, the kinetics of the transition could not be studied as a function of the density since the rotation weakens the transverse 2D confinement. We report hereafter a one-fold variation of the nucleation time of staggered states by varying the atomic density, and investigate experimentally and numerically the role of the renormalized tunneling rate and the modulation frequency on the transient out-of-equilibrium dynamics. We also explore the limit of the effective Hamiltonian approach in this context.

Our experiments have been realized on our rubidium-87 BEC machine that relies on a hybrid (magnetic and optical) trap \cite{PRL2016}.  It produces pure BECs of $10^5$ atoms in the lowest hyperfine level $F=1,m_F=-1$. The 1D optical lattice is generated by superimposing two counter-propagating laser beams at 1064 nm (lattice spacing $d=532$ nm) to the horizontal optical guide of the hybrid trap. The relative phase between the two arms of the lattice is controlled in time via synthesizers whose frequencies are imprinted on light using acousto-optic modulators. With a sinusoidal phase modulation of tunable amplitude $\varphi_0$, the atoms experience the potential 
$$
V(x)=\frac{1}{2}m\omega_{\rm ext}^2x^2-\frac{V_0}{2} \left[  1 + \cos\left(  \frac{2\pi x}{d} + 2\varphi_0\sin(2\pi \nu t) \right)   \right]
$$
where $\omega_{\rm ext}$ accounts for the longitudinal confinement of the hybrid trap along the guide axis.

To obtain a theoretical understanding of the behaviour of a BEC in the presence of such a modulated potential, we first perform a gauge transformation to the comoving frame in which the lattice is periodically tilted instead of shaken. A single-band approximation can be justified in this latter representation provided the amplitude of the lattice is sufficiently strong such that $V_0 > 4 \pi \varphi_0 m \nu^2 d^2$, which is the case in our experiments. The motion of the atoms along the lattice can then be modeled by means of a one-dimensional tight-binding Hamiltonian in which each well of the lattice is represented by one site. This Hamiltonian is constituted by site-dependent on-site energies that are essentially given by the longitudinal confinement of the trap, as well as by an approximately site-independent inter-site hopping matrix element $J$ that can be tuned by varying the strength and/or wavelength of the lattice.

As a result of the modulation, this inter-well tunneling rate $J$ is renormalized by a Bessel function according to
\begin{equation}
\bar J =  J \times J_0(2\pi\varphi_0 h\nu/E_L)
\label{jeff}
\end{equation}
where $E_L=h^2/(2md^2)$ is the lattice characteristic energy.  This result is readily derived from a one-body analysis \cite{holthaus} but turns out to remain valid in the presence of two-body interactions provided the latter as well as $J$ are significantly smaller than $h\nu$. A qualitative picture of the impact of this renormalization on the physics of the system can be worked out perturbatively with the expression for the energy of the lowest band using the Peierls substitution: $E_0(k)=-2\bar{J}\cos(kd)$. For $\bar{J}>0$ the minimum of the band is located at $k=0$, and the Fourier transform of the wave function consists in a comb of peaks centered about $k=0$ with a spacing $2\pi/d$. For $\bar{J}<0$, the minima are located on the border of the Brillouin zone at $k=\pm \pi/d$. 

When the sign of $\bar{J}$ is changed through phase modulation, the system is therefore put in a metastable state. 
While this would not affect the mean-field dynamics of a BEC in the presence of a translational invariant lattice, any deviation from perfect homogeneity in the condensate wavefunction or the lattice will give rise to a shrinking amplitude of the periodic condensate mode and to an exponentially increasing population of staggered modes at $k=\pm \pi/d$. The mechanism for the nucleation of the staggered states has been identified as a dynamical instability also referred to as a parametric instability triggered by interactions \cite{Lellouch17PRX}. Going beyond the mean-field description of the condensate, deviations from translational invariance are naturally present in form of quantum fluctuations by means of which the actual many-body dynamics of the quantum gas at hand can be conveniently modeled and understood. As a consequence, a quantum depletion of the condensate is expected to arise in the course of time evolution, followed by the formation of an incoherent, thermal-like, cloud of atoms within staggered modes near the border of the Brillouin zone \cite{rem_staggered}.

To obtain a qualitative understanding of this dynamical instability, we follow Ref.~\cite{Lellouch17PRX} and neglect the presence of the external longitudinal confinement for a moment. The system is then modeled by a homogeneous Bose-Hubbard Hamiltonian
\begin{equation}
  \hat{H} = - \bar{J}
  \sum_{l=-\infty}^\infty \left( \hat{b}_l^\dagger\hat{b}_{l+1} + 
    \hat{b}_{l+1}^\dagger \hat{b}_l \right)
  + \frac{U}{2}\sum_{l=-\infty}^\infty \hat{b}_l^\dagger\hat{b}_l^\dagger
  \hat{b}_l\hat{b}_l \label{eq:BH}
\end{equation}
where $\hat{b}_{l}$ denotes the annihilation operator associated with the Wannier function centered on the $l$-th well of the lattice and $U$ accounts for a site-independent two-body interaction strength. Assuming the presence of a perfectly homogeneous condensate at $t=0$ exhibiting $n$ atoms per site, we make the Bogoliubov ansatz \cite{CastinDum}
\begin{equation}
  \hat{b}_l(t) = \left[ \sqrt{n} + \sqrt{\frac{d}{2\pi}} \int_{-\pi/d}^{\pi/d}
    \hat{\Lambda}(k,t) e^{i l k d} dk \right] e^{-i \mu t / \hbar}
  \label{eq:Bog}
\end{equation}
where $\mu=- 2 \bar{J} +  n U$ is the chemical potential of the condensate. Linearizing the resulting equation for the thereby introduced de-excitation operators $\hat{\Lambda}(k,t)$ yields
\begin{eqnarray}
  i \hbar \frac{\partial}{\partial t} \hat{\Lambda}(k,t) & = & 
  2 \bar{J} ( 1 - \cos k d ) \hat{\Lambda}(k,t) \nonumber \\ 
  && + n U \left[ \hat{\Lambda}(k,t) + \hat{\Lambda}^\dagger(-k,t) \right]
\end{eqnarray}
whose solution evolves sinusoidally in time according to $\hat{\Lambda}(k,t) = [ \cos \Omega_k t - 2 i (\bar{J}/\Omega_k) ( 1 - \cos k d ) \sin \Omega_k t ] \hat{\Lambda}(k,0)- i (n U/\Omega_k) [\hat{\Lambda}(k,0) + \hat{\Lambda}^\dagger(-k,0)] \sin \Omega_k t$ with the Bogoliubov phonon frequencies
\begin{equation}
  \Omega_k = \frac{1}{\hbar} \sqrt{4 \bar{J}(1 - \cos  k d) 
    \left[\bar{J}(1 - \cos  k d) + n U \right]} \,.
\end{equation}
The time-dependent population of non-condensed modes can then be directly evaluated through
\begin{equation}
  \langle \hat{\Lambda}^\dagger(k,t) \hat{\Lambda}(k',t) \rangle
  = (nU)^2 \left(\frac{\sin \Omega_k t }{\Omega_k}\right)^2 
  \delta(k-k') \label{eq:popk} \,.
\end{equation}

A dynamical instability occurs when $\bar{J}$ becomes negative since $\Omega_k$ becomes imaginary for each quasiparticle mode that satisfies the relation $\bar{J}(1 - \cos k d) < 0 < \bar{J}(1 - \cos k d) + n U$, which according to Eq.~\eqref{eq:popk} implies that tiny initial populations of such modes experience an exponential growth with time. For $nU>-4\bar J$, this growth is most pronounced for the staggered Bogoliubov modes defined by $kd = \pm \pi$, which describe an antiperiodic Bloch function within the lattice; the population of these staggered modes grows with the Lyapunov exponent $\lambda=[-8\bar J(2\bar J +nU)]^{1/2}/\hbar$. This expression has stimulated our experiments described below, which explore how the nucleation time of the staggered states depend on the value of $\bar J$ and on the atomic density.

To allow for a more quantitative comparison with the experiment, we performed numerical simulations of the time evolution of the condensate by using the Truncated Wigner (TW) method \cite{TWBiblio}. This method approximately accounts for the effect of quantum fluctuations about the initial coherent state of the condensate. Applied to the Bose-Hubbard Hamiltonian \eqref{eq:BH}, it essentially amounts to sampling the time evolution of the quantum bosonic many-body state in terms of classical trajectories that evolve according to a discrete Gross-Pitaevskii equation.

For the practical implementation of the TW method, we take into account the presence of the longitudinal confinement of the hybrid trap, which gives rise to site-dependent on-site energy terms that have to be added to the Bose-Hubbard Hamiltonian \eqref{eq:BH}, as well as to a spatially inhomogeneous (Gaussian-like) condensate wavefunction which is calculated through imaginary time propagation. We furthermore incorporate the fact that the interaction energy within each site generally scales in a non-quadratic manner with the population of that site, due to the presence of a transverse Thomas-Fermi profile within the pancake-shaped lattice wells \cite{supp}. However, the presence of thermal fluctuations in the initial state (which can also be taken into account by the TW method) are neglected, i.e., we make the simplifying assumption that we start with a perfect BEC at $T=0$ K.

\begin{figure}[t]
\centering
\includegraphics[width=8cm]{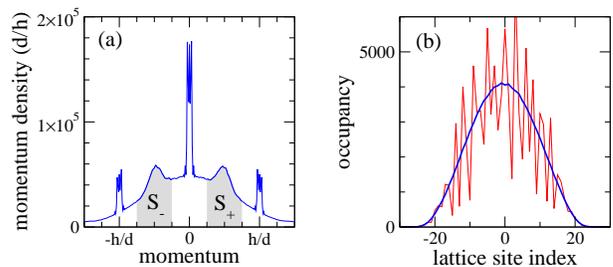}
\caption{(Color online) Numerically computed mean momentum density (a) and mean numbers of atoms on the lattice sites (b, blue line) at the evolution time $t=7.5$ ms in the presence of a shaking with the frequency $\nu = 1.5\,\mathrm{kHz}$ and the amplitude $\varphi_0=0.75\pi$, calculated with the TW method for a BEC containing $10^5$ atoms. The shaded areas in (a) correspond to the staggered mode populations $S_+$ and $S_-$. The light gray (red) line in panel (b) shows the occupancies that are obtained by computing a single TW trajectory. It indicates that the population of staggered states is accompanied by strong fluctuations of the lattice site occupancies.}
\label{figsim}
\end{figure}

The transition to staggered states is quantitatively characterized by the \emph{nucleation time} representing the instance at which the population of a periodic condensate state within the shaken lattice becomes less significant than the population of a staggered state with antiperiodic nature. To this end, we define by $P_0$ the population of the central condensate peak located at $k=0$, and by $S_\pm$ the population of staggered states centered about $\pm h/(2d)$ in momentum space. In the numerical simulations, $P_{0}$ and $S_\pm$ are determined by integrating the momentum-space density $\rho(k) = n(k) a_{||} \exp(-k^2 a_{||}^2) /(\pi^{1/2}d)$ within the intervals $-\pi/2d < k < \pi/2d$ and $-\pi/2d < k \mp \pi/d < \pi/2d$, respectively, where $n(k)$ is the $2\pi/d$-periodic momentum density that results from the TW simulation of the Bose-Hubbard dynamics [see the shaded areas in Fig.~\ref{figsim}(a)]. The nucleation time is then defined by the time beyond which $S_+ + S_-$ exceeds $P_0$. It corresponds to the instance at which the peaks at $k=\pm \pi/2d$ have the same accumulated contrast as the peak at $k=0$ in the experimental absorption images.

The first experimental study that we performed deals with the nucleation time of staggered states for different values of the renormalized tunneling rate $\bar J$. We specifically explored the dynamics at the border of the zone for which $\bar J$ becomes negative. The Lyapunov exponent predicts a divergence of the time required to populate the staggered states. We indeed observed a strong increase of the nucleation time when we approach an amplitude of modulation that corresponds to the zero of the Bessel function [see Fig.~\ref{figj0}(c)]. Good agreement is obtained with TW computations of the nucleation time indicated by the solid (dashed) lines in Fig.~\ref{figj0}(b), which were conducted according to the above prescriptions assuming the presence of $10^5$ ($5\times10^4$) atoms in the condensate. As these simulations were performed at $T=0$, thermal fluctuations seem to play a minor role in this set of data for which the temperature was indeed so low that it could not be measured by time-of-flight methods.

The evolution of the atomic gas in momentum space is illustrated in Fig.~\ref{figj0}(a) which shows a sequence of time-of-flight absorption images taken after various evolution times. We clearly see the transition from an initially coherent BEC, characterized by a sharp central peak at $k=0$ and by two side peaks at $k = \pm 2\pi/d$, to an incoherent thermal cloud that oscillates between two maxima at $k = \pm \pi/d$. These latter peaks are significantly broader than the initial condensate peaks, which is indicative of an effective increase of the temperature as a consequence of the dynamical instability mechanism \cite{Lellouch17PRX,strater,Huveneers,Bloch2017}. This is also observed in the TW simulations, as is seen in Fig.~\ref{figsim}(a) which shows a snapshot of the numerically computed momentum distribution of the atomic gas after the nucleation process. Note that the sharp coherent BEC peaks at $k=0$ and $\pm \pi/d$ are still present in the numerical simulations, in contrast to the experiment where they are nearly completely washed out after the nucleation. We attribute this to the fact that our TW approach is based on a single-band approximation and does not account for the effect of quantum fluctuations in the transverse degrees of freedom within the lattice, which are expected to provide further significant contributions to the quantum depletion of the condensate \cite{BKT}.

The TW approach allows one to obtain complementary insight into the nature of the staggered states that would not be easily accessible in the experiment. This concerns, in particular, the behaviour of lattice site occupancies. While their mean values $\langle\hat{n}_l\rangle = \langle \hat{b}_l^\dagger \hat{b}_l \rangle$ do not display any notable feature in the course of time evolution, their rms widths $(\langle\hat{n}_l^2 \rangle - \langle\hat{n}_l\rangle^2)^{1/2}$ dramatically increase as soon as staggered states become significantly populated. This is illustrated in Fig.~\ref{figsim}(b) where the mean lattice site occupancies (averaged over 10000 trajectories) are plotted together with the occupancies that were obtained from a single trajectory (shown in red), at an evolution time $t = 7.5$ ms that exceeds the nucleation process.

The occurrence of pronounced spatial fluctuations is strongly reminiscent of \textit{gap solitons} \cite{gapsolitons} and indicates that the formation of staggered states in momentum space is accompanied by the generation of solitons. This interpretation is consistent with the Bogoliubov mode analysis in a 1D shaken optical lattice of Ref.~\cite{Lellouch17PRX}. It is further confirmed by the fact that the spatial extension of the density peaks is indeed on the order of the theoretical prediction $\sigma \simeq 2.6 d [2|\bar{J}|/(n_{\rm max} U)]^{1/2} \simeq 0.7 d$ for the full width at half maximum of a gap soliton according to Ref.~\cite{gapsolitons} with $n_{\rm max}\simeq 6000$ the maximal occupancy within a lattice site. We should keep in mind, however, that in the actual (three-dimensional) experimental context, these gap solitons are expected to quickly disintegrate into vortices and vortex rings through the snake instability \cite{snake}. This effect is not accounted for in our numerical simulations which are based on a one-dimensional approximation for the optical lattice.

\begin{figure}[t]
\centering
\includegraphics[width=8cm]{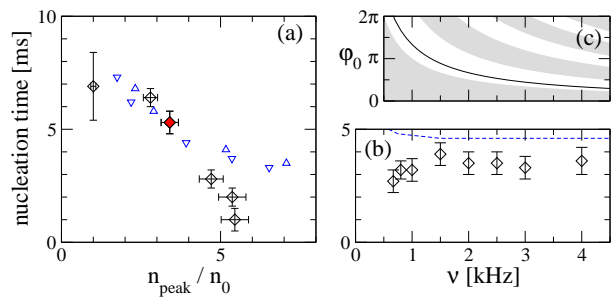}
\caption{(Color online) (a) Nucleation time as a function of the peak atomic density for an optical lattice of depth of $2.6E_L$ ($n_0 \simeq 10^{13}$ at.cm$^{-3}$), a driving frequency of 1.5 kHz and an amplitude of modulation of $\varphi_0=0.75\pi$. Upper (lower) triangles show the results of numerical TW simulations taking into account $N = 10^5$ ($5\times 10^4$) atoms in the condensate and using accurate estimates for the longitudinal and transverse trapping frequencies of the experiments. The (red) filled diamond corresponds to the time-of-flight images shown in Fig.~\ref{figj0}(a). (b) Nucleation time as a function of the shaking frequency $\nu$ where the shaking amplitude $\varphi_0$ is adapted such that $\nu\varphi_0$ (and hence also $\bar{J}$) is kept constant [see panel (c); shaded areas show parameter regions where the effective tunneling matrix element \eqref{jeff} is positive]. TW simulations (dashed blue line) predict a convergence towards the nucleation time that would be obtained within the time-averaged Bose-Hubbard model \eqref{eq:BH}. Performed at $T=0$ K, they provide an upper bound on the nucleation time since they do not take into account the role of thermal fluctuations.}
\label{figkinetics}
\end{figure} 

To explore the dependence of the nucleation time on the atomic density, we performed experiments where we drastically diminished the intensity of the vertical beam of the crossed dipole trap after the BEC production. In this manner, we could vary the atomic peak density between $5.5 \times 10^{13}$ at.cm$^{-3}$ and $10^{13}$ at.cm$^{-3}$ \cite{calibration} and observe a one fold increase of the nucleation time. As shown in Fig.~\ref{figkinetics}(a), the experimental results are in fairly good agreement with TW simulations, except for high peak densities $n_{\rm peak} \gtrsim 5\times10^{13}$ at.cm$^{-3}$. Indeed, to get such high densities we compress adiabatically the trap and therefore increase the temperature. Our results correspondingly suggest that thermal fluctuations are playing here a more important role in the nucleation process.

Finally, we investigate in Fig.~\ref{figkinetics}(b) the scaling of the nucleation time with the driving frequency $\nu$ where we adapt the shaking amplitude $\varphi_0 = (0.889 \pi / \nu) \times 1.5\,\mathrm{kHz}$ such that the argument of the Bessel function is kept constant according to Eq.~\eqref{jeff} yielding $\bar{J} \simeq - 0.33 J$. As we have the same time-averaged Bose-Hubbard Hamiltonian \eqref{eq:BH} for all $\nu$, the nucleation time is found to vary only rather weakly with the driving frequency. Note, however, that this behaviour is expected to change for $\nu \gtrsim V_0 / ( 4 \pi \varphi_0 \nu m d^2) \simeq 6.5\,\mathrm{kHz}$ where the single-band approximation can no longer be justified within the comoving frame. Indeed, a significant population of higher bands is expected to occur in this latter high-frequency regime as a consequence of near-resonant transitions, which in turn should lead to a drastic decrease of the nucleation time. This is exemplified in an experimental study performed for $\nu = 14\,\mathrm{kHz}$ and $\varphi_0=0.028\pi$ (all other parameters were chosen as usual) for which a transition to staggered states is found after a few ms, despite the fact that $\bar{J} > 0$ for this combination of parameters \cite{supp}. We attribute this instability of the condensate to a near-resonant transition from the ground band to the first excited band in the lattice. The latter exhibits a maximum at $k=0$ in the Brillouin zone, which gives rise to a negative inter-site hopping matrix element.

In summary, we presented an experimental and theoretical study of the formation of staggered states within a BEC that is subjected to a periodically modulated optical lattice. Our measurements are in good agreement with numerical simulations using the TW method at $T=0$ K which are based on a single-band description of the lattice, except for high densities where we have evidence that thermal fluctuations are playing an important role. The present experimental setting can therefore be exploited in order to yield a quantitative diagnostic tool for determining interaction and fluctuation effects in BECs. Further studies in the high-frequency regime will be useful and interesting in order to explore the interplay with near-resonant inter-band transitions in more detail.

We thank J. Dalibard, N. Goldman, and M. Oberthaler for useful discussions. This work was supported by Programme Investissements d'Avenir under the program ANR-11-IDEX-0002-02, reference ANR-10-LABX-0037-NEXT, and by the Programme Hubert Curien 2016.

\newpage

\onecolumngrid

\begin{center}

\textbf{\large Supplementary material for \\[0.2em]
'Out-of-equilibrium dynamics of a Bose Einstein condensate in a periodically driven band system'} \\[1em]

E. Michon$^1$, C. Cabrera-Guti\'errez$^1$, A. Fortun$^1$, M. Berger$^1$, M. Arnal$^1$, \\
V. Brunaud$^1$, J. Billy$^1$, C. Petitjean$^2$, P. Schlagheck$^2$, D. Gu\'ery-Odelin$^1$\\
$^1$ \textit{Universit\'e de Toulouse ; UPS ; Laboratoire Collisions, Agr\'egats, R\'eactivit\'e, IRSAMC ; F-31062 Toulouse, France\\
CNRS ; UMR 5589 ; F-31062 Toulouse, France and}\\
$^2$ \textit{D\'epartement de Physique, CESAM research unit, University of Liege, 4000 Liege, Belgium}\\[0.5em]
\end{center}

\twocolumngrid

In this supplementary material, we provide some technical details on the implementation of the Truncated Wigner method which was used to perform the numerical simulations. We furthermore provide a reminder on our lattice depth calibration technique  \cite{fortun}. We finally show results of phase modulation experiments performed in an extended range of parameters in both modulation frequency and lattice depth with respect to the experiments presented in the main article and discuss the nucleation of staggered states in each case.

\section{Implementation of the Truncated Wigner method}

The starting point for the implementation of the Truncated Wigner method is the effective many-body Bose-Hubbard Hamiltonian
\begin{eqnarray}
  \hat{H}(t) & = & - J \sum_{l = -\infty}^{\infty} 
  \left[ \hat{b}_l^\dagger \hat{b}_{l+1} e^{-i \theta(t)} +
    \hat{b}_{l+1}^\dagger \hat{b}_l e^{i \theta(t)} \right] \nonumber \\ &&
  + \sum_{l = -\infty}^{\infty} \left[ V_l \hat{b}_l^\dagger \hat{b}_l
    + \frac{U_l}{2} \hat{b}_l^\dagger \hat{b}_l^\dagger \hat{b}_l \hat{b}_l \right]
  \label{eq:bh}
\end{eqnarray}
which describes the dynamics of a Bose-Einstein condensate within a periodically shaken lattice. Here we employ a single-band approximation within the reference frame that is comoving with the shaken lattice. The shaking is incorporated by means of a periodically time-dependent Peierls phase 
\begin{equation}
  \theta(t) = 2 \pi \varphi_0 \frac{h \nu}{E_L} \cos(2\pi\nu t)
\end{equation}
within the inter-site hopping matrix elements of the lattice. The on-site energies
\begin{equation}
  V_l = \frac{1}{2} m \omega_{||}^2 d^2 l^2
\end{equation}
account for the presence of the longitudinal harmonic confinement with the oscillation frequency $\omega_{||}$.

For lattice strengths $V_0 = s E_L$ with $s\gtrsim 2$, we can approximately represent the Wannier function within the $l$th well by the Gaussian
\begin{equation}
  \Phi_l(x) = \frac{1}{\sqrt{\pi^{1/2} a_0}} e^{- (x-ld)^2 / (2 a_0^2)} \label{eq:wannier}
\end{equation}
where 
\begin{equation}
  a_0 = \sqrt{\frac{\hbar}{m \omega_0}} = \frac{\sqrt{2 s^{-1/2}}}{k}
\end{equation}
is the oscillator length associated with the frequency $\omega_0 = s^{1/2} E_L / \hbar$ of oscillations within each well of the lattice.
The effective hopping parameter between adjacent wells can be determined
from a semiclassical Wentzel-Kramers-Brillouin (WKB) ansatz \cite{landau,Garg}.
We obtain
\begin{equation}
  J = \frac{\hbar \omega_0}{\sqrt{e\pi}} e^{-\sigma}
\end{equation}
with the dimensionless imaginary action
\begin{eqnarray}
  \sigma & = & \sqrt{2 s} \int_{\arccos \eta}^\pi \sqrt{\eta - \cos u} \, du 
  \nonumber \\
  & = & \sqrt{8 s( \eta + 1)} E\left[ \frac{\pi - \arccos \eta}{2}, 
    \sqrt{\frac{2}{\eta + 1}} \right]\,,
\end{eqnarray}
where we introduce the parameter
\begin{equation}
  \eta = \exp\left(-\frac{1}{2 s^{1/2}}\right) - \frac{1}{2 s^{1/2}}
\end{equation}
and the incomplete elliptic integral of the second kind
$E(\varphi,k) = \int_0^\varphi\sqrt{1 - k^2 \sin^2 \theta} d\theta$.
Note that this WKB ansatz is based on the approximate expression
$E_0 = (1 - \eta) s^{1/2} \hbar \omega_0/2$
for the ground state energy within each well, which is obtained from the expression \eqref{eq:wannier} for the Wannier function using first-order perturbation theory.

A complication is introduced by the fact that the lattice wells in the experiment are not truly one-dimensional but rather exhibit a pancake shape, as the transverse confinement frequency $\omega_\perp$ of the trap is comparable to the longitudinal one $\omega_{||}$ and hence much larger than $\omega_0$. For the total atom numbers at hand, the BEC exhibits a parabolic Thomas-Fermi profile in its transverse density distribution within lattice wells that are located near the center of the trap. The interaction energy within such a well therefore scales in a non-quadratic manner (namely $\propto n_l^{3/2}$) with the occupancy $n_l$ of that well, which implies that the effective on-site interaction parameters $U_l$ appearing in the Bose-Hubbard Hamiltonian \eqref{eq:bh} scale as $n_l^{-1/2}$ with the occupancies of the corresponding sites. For weakly populated lattice sites that are located at the edge of the atomic cloud, on the other hand, we can justify the perturbative approximation $g \simeq 2 \hbar \omega_\perp a_s$ \cite{olshanii} for the effective one-dimensional interaction strength with $a_s \simeq 5.3\times 10^{-9}\,\mathrm{m}$ the $s$-wave scattering length of $^{87}$Rb atoms, which yields $U_l \simeq 2 \hbar \omega_\perp a_S / (\sqrt{2\pi} a_0)$ for those sites.

In order to simultaneously account for weakly and strongly occupied sites of the lattice, we employ a heuristic interpolation formula
\begin{equation}
  U_l = \frac{2 \hbar \omega_\perp a_s / ( \sqrt{2\pi} a_0 )}{\sqrt{1 + 
      4 n_l a_s / ( \sqrt{2\pi} a_0 )}} \label{eq:Ul}
\end{equation}
between the perturbative and the Thomas-Fermi regime \cite{Paul}. We furthermore make the simplifying assumption that the lattice site occupancies $n_l$ remain fairly constant and are not substantially altered in the course of time evolution. While this constraint appears to be well respected on average, significant fluctuations of the lattice site occupancies about their average values may nevertheless arise if the condensate undergoes a transition to a staggered state [as seen in Fig.~2(b) of the main article].

Having thereby determined all relevant parameters of the Bose-Hubbard system
 \eqref{eq:bh}, we can then derive the discrete nonlinear Schr\"odinger equation
\begin{eqnarray}
  i \hbar \frac{d}{dt} \psi_l(t) & = & 
  - J \left[\psi_{l+1}(t) e^{i \theta(t)} + \psi_{l-1}(t) e^{-i \theta(t)} \right]
  \nonumber \\ &&
  + V_l \psi_l(t) + U_l \left[ |\psi_l(t)|^2 - 1 \right] \psi_l(t) \label{eq:TW}
\end{eqnarray}
by means of which the classical fields $\psi_l$ that sample the quantum many-body state of the system have to be evolved with time in the framework of the Truncated Wigner method. For the initial state we assume the presence of a perfect BEC within the optical lattice, which is in an idealized manner described by a coherent state in the classical field space. This leads to the choice $\psi_l(0) = \phi_l + \chi_l$ for the initial value of $\psi_l$ where $\phi_l$ corresponds to the condensate wavefunction within the (unshaken) lattice at $t=0$. In practice, $\phi_l$ is determined by imaginary-time propagation of the Gross-Pitaveskii equation that describes the condensate at $t=0$ [which is nearly identical with Eq.~\eqref{eq:TW}, except that it exhibits real hopping and the usual $U_l |\psi_l|^2 \psi_l$ interaction term]. This thereby yields $n_l = |\phi_l|^2$ for the determination of the on-site interaction strengths according to Eq.~\eqref{eq:Ul}. $\chi_l$ is a complex random number drawn from a Gaussian probability distribution which is centered about the origin in the complex plane and yields the variance $\overline{|\chi_l|^2} = 1/2$. Effects due to the presence of finite temperature and initial quantum depletion within the atomic cloud are therefore neglected in this study.

Expectation values of one-body operators are then evaluated according to the
prescription
\begin{equation}
  \langle \hat{b}_l^\dagger \hat{b}_{l'} + \hat{b}_{l'} \hat{b}_l^\dagger \rangle_t
  = 2 \overline{\psi_l^*(t) \psi_{l'}(t)}
\end{equation}
where $\overline{\psi_l^*(t) \psi_{l'}(t)}$ denotes the statistical average (over Truncated Wigner trajectories) of the expression $\psi_l^*(t) \psi_{l'}(t)$. This yields in particular $\langle \hat{b}_l^\dagger \hat{b}_l \rangle_t = \overline{|\psi_l(t)|^2} - 1/2$ for the average occupancy of the site $l$ at time $t$. A similar subtraction of a ``half-particle'' is needed in order to determine the average distribution of atoms in momentum space.

\section{Lattice depth calibration}

\begin{figure}[h!]
\centering
\includegraphics[width=7cm]{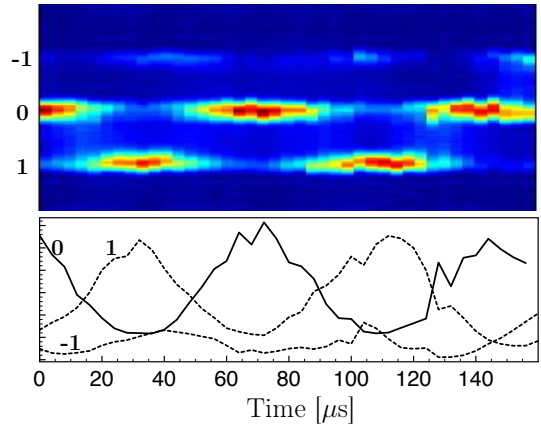}
\caption{Lattice depth calibration using the oscillatory motion of atoms within the lattice after a sudden phase shift of the lattice of $\theta_0=90^{\circ}$ at $t=0$. The measured period is 153 $\mu$s, corresponding to a lattice depth of 1.9 $E_L$. Top: absorption images for various holding time in the shifted lattice (followed by 25 ms time of free flight in the absence of the lattice). Bottom: time evolution of the populations of the different orders of momentum $p=(0, \pm 1) h/d$ extracted from the absorption images.
}
\label{fig1}
\end{figure} 

We perform a precise calibration of the lattice depth using the out-of-equilibrium dynamics of a chain of BECs in an optical lattice following the method we demonstrated in \cite{fortun}. To do so, we first load adiabatically a BEC in the lattice, creating a chain of BECs trapped at the bottom of the lattice sites. At $t=0$, we suddently shift the phase of the lattice, which triggers the center-of-mass motion of the atomic wave packets in each well (shift of $\theta_0=90^{\circ}$ here). After a given holding time in the shifted lattice, we perform a 25 ms time-of-flight. The experimental absorption images (see Fig.\ref{fig1}) show the interferences of the different wave packets located in each lattice site. The interference figure is centered on the central interference peak (0th order) corrresponding to wave packets that were released while being at rest in the shifted lattice, i.e. at the turning points of the oscillatory motion. We observe oscillations of the population in the 0th order from which we extract the period of the center-of-mass motion of the atomic wave packets. This period can be directly related to the depth of the optical lattice. Indeed, the intrasite dipole mode is coupled to a two phonon transition between the ground state band and the second excited band at $k=0$:
\begin{equation}
T_0=\frac{2h}{E_3(k=0)-E_1(k=0)}.
\end{equation}
In this way, we find a lattice of depth 1.9 $E_L$ corresponding to the measured period  $T_0$=153 $\mu$s. 
Interestingly, we have shown that this kind of measurement is robust against the value of the phase shift of excitation $\theta_0$, the atom-atom interaction strength, and the external confinement superimposed to the lattice. 
For these values of the phase shift and the lattice depth, tunneling plays an important role in the wave packet dynamics insofar as an important fraction of the wave packets tunnels to the neighboring lattice sites. It explains for instance the population of the -1 order for an evolution time in the shifted lattice of 40 $\mu$s but also more generally the asymmetry of the populations in the $\pm$ 1 orders. We have studied in details such effect in Ref.~\cite{fortun}.

\section{Parameter ranges for the nucleation of staggered states}

\subsection{Frequency range}

\begin{figure*}[h!]
\centering
\includegraphics[width=0.7\textwidth]{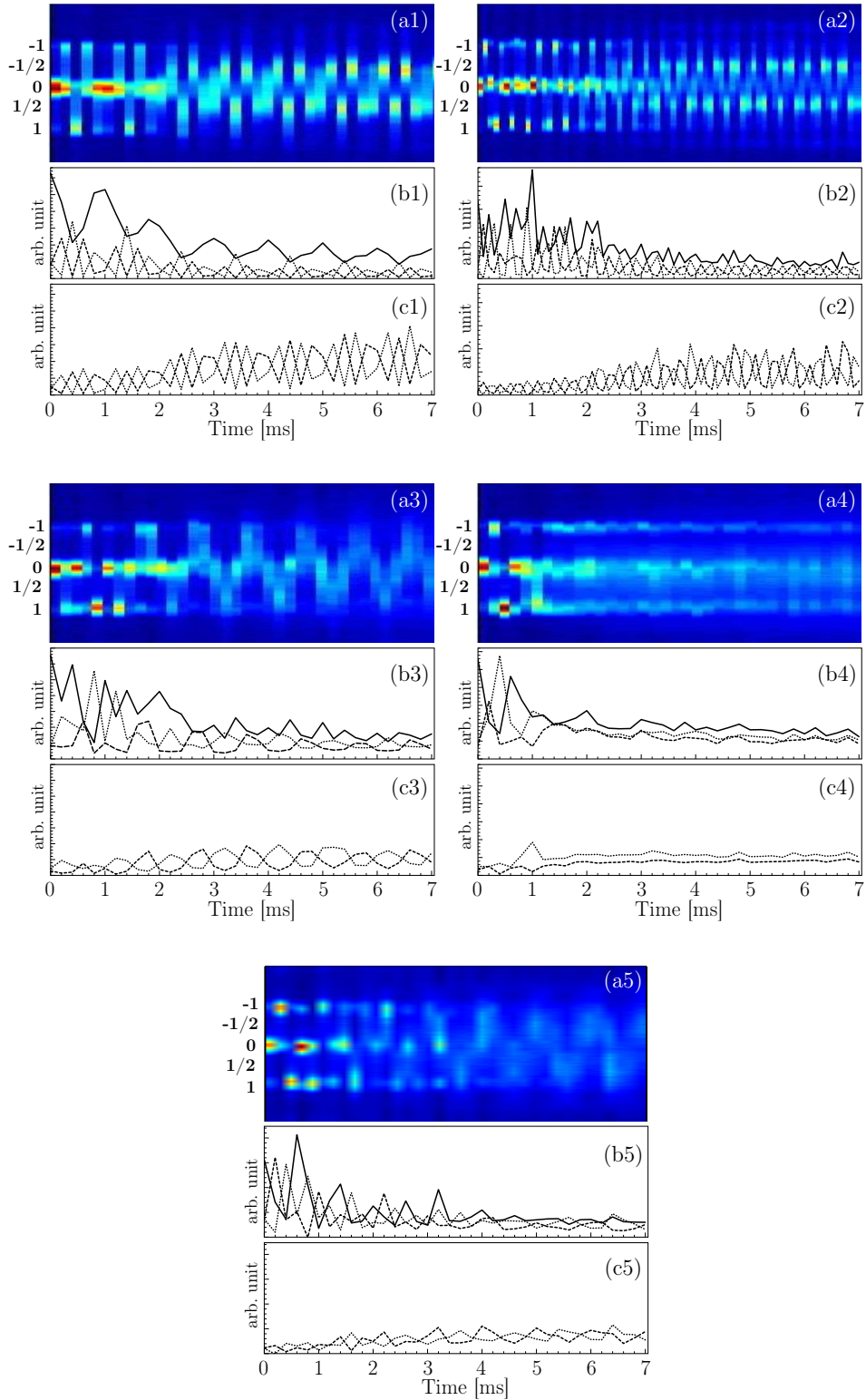}
\caption{Nucleation of the staggered states for different modulation frequency $\nu$. For (1-4), the modulation frequencies are below the center-of-mass oscillation frequency ($\nu_c = 6.55$ kHz, corresponding to a lattice depth of 1.9 $E_L$). The product $\varphi_0 \nu$ with $\varphi_0$ the amplitude of modulation is kept constant, corresponding to a negative renormalized tunneling rate $\bar J$. The (amplitude,frequency) couple takes the following values (1) (0.67$\pi$, 2~kHz), (2) (0.44$\pi$, 3~kHz), (3) (0.33$\pi$, 4~kHz) and (4) (0.27$\pi$, 5~kHz). For (5) the modulation frequency ($\nu = 14$ kHz) is above the center-of-mass oscillation frequency, in this case $\nu_c = 10.6$ kHz (lattice of depth 3.7 $E_L$) and the amplitude of modulation ($\varphi_0=0.028\pi$) such that one would get a positive $\bar J$ within the single-band approximation. The approximation breaks down in this frequency domain and we could nucleate staggered states. For every set of values are represented: (a) absorption images obtained for various evolution times followed by 25 ms time-of-flight, where the time step is set to 100$\mu$s for the measurement (2) and 200$\mu$s for the other measurements; (b,c) time evolution of the population in the different orders of momentum $p=(0,\pm 1/2, \pm 1) h/d$ extracted from the absorption images, with the orders (b) 0 (solid), +1 (dashed), -1 (dotted) and (c) +1/2 (dashed), -1/2 (dotted).}
\label{fig2}
\end{figure*}

We extend the study of the effect of the modulation frequency $\nu$ on the nucleation of staggered states performed in the main article to determine the frequency range in which such states appear. We first use a lattice of depth 1.9 $E_L$, characterized by a center-of-mass oscillation of frequency $\nu_c = 6.55$ kHz (see above) and perform phase modulation experiments at different modulation frequencies below $\nu_c$ [see Fig.\ref{fig2} (1-4)]. We maintain the product $\varphi_0 \nu$ constant with $\varphi_0$ the amplitude of modulation, yielding an effective tunneling rate $\bar J$ that is constant and negative. The argument of the Bessel function [see equation (1) of the main article] is chosen equal to $3.24$.

At the lowest modulation frequencies (2 and 3 kHz), we clearly observe the nucleation of staggered states with the population of staggered modes at momentum $p=\pm h/2d$,  whereas for larger modulation frequencies ($\nu$ = 5 kHz), we do not significantly populate the staggered modes [see Fig.\ref{fig2}(c4)]. More specifically, defining by $\nu_c$ the center-of-mass oscillation frequency within a lattice well, we observe the population of staggered modes for modulation frequencies up to typically $\nu_c/2$, meaning for modulation frequencies that are not too close to excitations frequencies towards the excited bands. The experiments presented in the main article on the effect of the modulation frequency (see Fig. 3 of the main article) were performed with a lattice of depth $2.6 E_L$, corresponding to a center-of-mass oscillation frequency $\nu_c=8$ kHz. As those experiments were performed for modulation frequencies up to 4 kHz, we could see the population of the staggered modes for each chosen modulation frequency.

Additionally, we perform modulation experiments in a lattice of larger depth, characterized by a center-of-mass ocillation of frequency $\nu_c = 10.6 $ kHz. Intriguingly, we could observe the nucleation of staggered states [see Fig.\ref{fig2} (5)] for a frequency of modulation ($\nu = 14$ kHz) larger than $\nu_c$ and a very low amplitude of modulation ($\varphi_0=0.028\pi$). For such parameters, the argument of the Bessel function is equal to 0.94 and would correspond to a positive effective tunneling rate in a single-band theory. However, the modulation frequency is close to the one-phonon line from the lowest to the first excited band. Thus, the theoretical treatment used in the main article, which is based on a single-band approximation, is no longer applicable. The study of the appearance of staggered states in this frequency regime requires a more involved treatment taking into account higher bands.

 \subsection{Lattice depth range}

\begin{figure*}[h!]
\centering
\includegraphics[width=0.99\textwidth]{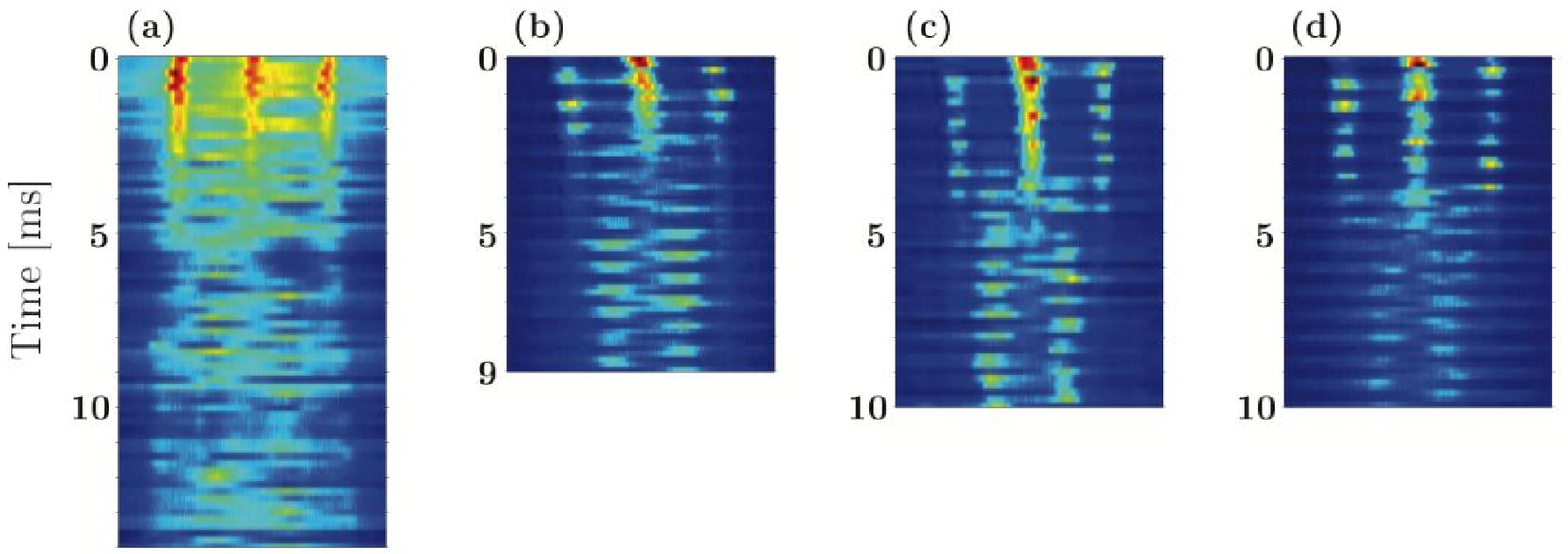}
\caption{Nucleation of the staggered states for different lattice depths: (a) 1.2 $E_L$, (b) 1.8 $E_L$, (c) 2.6 $E_L$ and (d) 3.2 $E_L$. The absorption images are obtained for various evolution times followed by 25 ms time-of-flight. The amplitude $\varphi_0$ and frequency $\nu$ of modulation have been chosen such to keep the argument of the Bessel fonction (see equation (1) of the main article) equal to 3.24: (a) $\varphi_0=1.67\pi$ and $\nu=800$ Hz (b-d) $\varphi_0=0.89\pi$ and $\nu=1.5$ kHz.}
\label{fig3}
\end{figure*}

We perform phase modulation experiments for different lattice depths ranging from $1.2 E_L$ to $3.2 E_L$ and observe the nucleation of staggered states in this range of lattice depth (see Fig.\ref{fig3}). The product $\varphi_0 \times\nu$ is kept constant and corresponds to a negative effective tunneling rate $\bar J$. The dynamics for the population of staggered modes look similar to the dynamics shown in the main article (see Fig. 1 in the main article), but for the lattice depth $1.2 E_L$. For the lattice depth $1.2 E_L$, the measurement is performed at a lower modulation frequency and as a consequence with a larger amplitude of modulation. For such an amplitude of modulation, the associated classical phase space starts to exhibit chaotic zones, which is probably responsible for the disrupted aspect of the dynamics observed in this case [see Fig.\ref{fig3}(a)].

\begin{thebibliography}{99}
\bibitem{rem_staggered}
We should note that a \emph{qualitatively} similar transition from a periodic to a staggered state can also arise on the level of the mean-field dynamics of the condensate described by the Gross-Pitaevskii equation, namely if the spatial homogeneity of the lattice is perturbed by the presence of an additional confinement. For the experimental parameters at hand, however, this perfectly coherent effect is entirely overshadowed by the depletion of the condensate arising due to the presence of quantum fluctuations.
\end{thebibliography}
\end{document}